\documentclass[review]{elsarticle}

\usepackage{lineno,hyperref}
\usepackage{amsmath,amsfonts}
\usepackage{xspace}
\usepackage{tikz}  
\usepackage{pgfplots}
\usepackage{subfigure}
\usepackage{stfloats}
\usepackage{array}
\usetikzlibrary{shapes}
\usetikzlibrary{matrix}
\usetikzlibrary{patterns}
\usetikzlibrary{backgrounds}
\usepackage{pgfplots}
\usepackage{xspace}
\usepackage{booktabs}
\modulolinenumbers[5]

\journal{Journal of \LaTeX\ Templates}
\begin{document}

\begin{frontmatter}

\title{MixNN: A design for protecting deep learning models}
\author{Chao Liu}
\author{Hao Chen}
\author{Yusen Wu}
\author{Rui Jin}
\address{University of Maryland, Baltimore County
1000 Hilltop Circle, Baltimore, MD 21250}
\cortext[mycorrespondingauthor]{Corresponding author: Rui Jin}
\cortext[mycorrespondingauthor]{Hao Chen and Rui Jin made equal contributions}

\begin{abstract}
This work proposes a novel design, called MixNN, for protecting deep learning model structure and parameters both in the training phase. The layers in a deep learning model of MixNN are fully decentralized. 
Compared with the existing most popular two-server style structure, MixNN hides communication address, layer parameters and operations, and forward as well as backward message flows among non-adjacent layers so that it can defend against collusion attacks. MixNN has following advantages: \romannumeral1)~an adversary cannot fully control all layers of a model including the structure and parameters, \romannumeral2)~even some layers may collude but they cannot tamper with other honest layers, \romannumeral3)~model privacy is preserved both in the training phase and inference phase. \romannumeral4)~MixNN software is easy to be deployed and evaluated. In one classification experiment, we compared a neural network deployed in a virtual machine with the same one using the MixNN design on the AWS EC2. The result shows that our MixNN retains less than 0.001 difference in terms of classification accuracy, while the whole running time of MixNN is about 7.5 times slower than the one running on a single virtual machine. 
\end{abstract}

\begin{keyword}
\texttt{Deep learning, distributed system}
\end{keyword}

\end{frontmatter}

\section{Introduction}
Privacy
protection of deep learning (DL) models is important for guarding commercial and intellectual property, for example, a financial company may hold a private model which can facilitate stock investment; leakage of such model causes huge loss \cite{survey21}. Protecting DL models contains both model structure and model parameters. 

Privacy concerns of a DL model occur when deploying a model locally or on a cloud server. Deploying DL models in a local machine very probably leaks all the details of models to hackers and malicious colleagues. Meanwhile, machine learning as a service (MLaaS), such as, Amazon Machine Learning services \cite{AmazonML}, Microsoft Azure Machine Learning \cite{MicrosoftML}, Google AI Platform \cite{googleML}, and IBM Watson Machine Learning \cite{IBMML}, enables customers to use powerful DL tools and computation resources by deploying their DL models on the cloud. Even though private data from a client can be protected by using Intel SGX \cite{SGXml} or homomorphic encryption (HE) \cite{HEml}, an obvious problem is that the model may be still opened to some providers. Another issue is that the malicious cloud controller can easily steal the DL model by checking the codes, or obtain a close model by generating querying results and searching the DL model space \cite{CacheTelepathy, StealingMachine}. As a consequence, model privacy is not preserved.

Based on above issues, we propose the MixNN to explain: \textit{how to deploy a DL model on a powerful AI platform while protecting model privacy.}

Inspired by split learning \cite{VGSR2018}, MixNN distributes each layer in a DL model on one server. We assume that an adversary cannot control most of the layers. In this way, it prevents an adversary from acquiring the whole model structure and model parameters. We will discuss a special scenario where an adversary controls both the first layer and the last layer in MixNN in the Sec. \ref{medel_privacy}. We consider MixNN only on a cascade topology where forward and backward propagation must be proceeded through consecutive layers.

 One problem in this design is that one server controlled by an adversary in this cascade can figure out who the other servers are via decoding the message flow, then this malicious server could tamper with other honest servers by rewarding them (\emph{e.g.}, bitcoin) for exchanging  information. Another situation is that mutually acknowledged servers can cooperate together to disclose the sensitive model and data for common interest. If we do not have a way to hide the detailed physical address among these servers, an adversary can easily acquire the information by passively listening to the channel among these mutually acknowledged servers. Thus, the model structure and parameters of these layers on these malicious servers are exposed.

To tackle this issue, we adapt a novel method. In MixNN \cite{mixnetworkWIKI}, each message is encrypted to each layer using public key cryptography. Each layer strips off its own layer of encryption to reveal where to send the message next. In Fig. \ref{fig_model}, we take a communication process from layer 1 to layer $n$ as an example, the layer 2 only knows that it receives messages from layer 1. It then uses its secret key to decrypt the message flow from layer 1 and gets layer 3's physical address, and finally it sends the message to layer 3. Layer 2 has no knowledge about the address from layer 4 to layer $n$ assuming no failures occur. In MixNN, we use this approach to pack the message on the client slide to hide the detailed communication process among non-adjacent layers. In summary, even some layers are controlled by an adversary but it's hard for it to locate other honest layers in the network. 

Training a DL model, however, is different from using mix networks to realize anonymous communication. The $i$th layer in the model should compute the input set $\mathcal{Z}_i$ for the next layer in the forward propagation, and update the parameter set $\mathcal{W}_i$ in this layer by gradient descent during the backward propagation. In MixNN, layer $i$ decrypts the $\mathcal{Z}_{i-1}$ from layer $i-1$ to proceed DL operations. It also decrypts the intermediate gradient $\frac{\partial{l}}{\partial{z}_{i}}$, where $l$ represents the training loss and ${z}_{i}\in\mathcal{Z}_i$, from layer $i+1$ to update $\mathcal{W}_i$, and prepares the intermediate gradient $\frac{\partial{l}}{\partial{z}_{i-1}}$ for layer $i-1$. We use the chain rule in updating parameters and preparing gradients. Finally, layer $i$ uses the next layer's public key to encrypt forward or backward propagation message. Our MixNN fits a cascade topology which is one of the structures \cite{surveyMixnet} (cascade topology and free-routing topology) in mix networks. An adversary who is listening to the channel cannot learn both the DL computation result in layer $i$ and any information of DL parameters and operations in other non-adjacent layers.
\begin{figure*}
  \centering
  \includegraphics[scale=0.5]{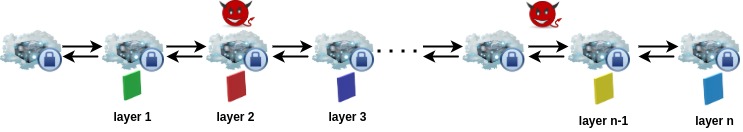}
  \caption{MixNN's design overview.}
  \label{fig_model}
\end{figure*}

In summary, MixNN not only distributes layers in a DL model on different servers so that an adversary can hardly control the whole model structure and parameters, but also hides communication address, layer parameters and operations, and forward and backward message flow among non-adjacent layers. Fig. \ref{fig_model} shows the MixNN's design. An adversary can control the server and passively listen to the channel between two servers. The square with color means a layer in a DL model and this colored layer is deployed on one server. All layers are constructed to one DL model. Some servers which do not have a layer on it are dummy servers and we will explain this in the Sec. \ref{system setup}.

\subsection{Contributions}
\label{subsec:contrib}
\begin{itemize}
    \item MixNN is a novel design for protecting model structure and parameters. Compared with previous works, MixNN decentralizes layers in a DL model on different servers instead of two parties (some layers on the client side and rest on the server side). This distributed method decreases the possibility which an adversary controls the whole structure and parameters of a model. 
    \item MixNN is the first design to use the ideas from mix networks for hiding real "identities" of non-adjacent layers in a cascade topology in DL structures. In this design, MixNN actually isolates every layer in a black box. An adversary can hold some black boxes and get parameters and operations but they cannot locate and control all of them. When transferring message layer by layer, each layer encrypts forward and backward propagation message to avoid leaking model information to the adversary who is passively listening to the channel.
    \item We provide a detailed description for deploying MixNN. It contains how to decentralize layers and how to use the way from mix networks to pack message in different DL phases. There are four phases in MixNN: model initialization, forward propagation phase, backward propagation phase, and testing phase, separately. The implementation follows the description of MixNN. Compared with a same neural network deployed in a single server on AWS EC2, we show that the MixNN has less than 0.001 difference in terms of classification accuracy, while the whole running time is about 7.5 times slower than the one run in a single virtual machine.

\end{itemize}

\subsection{Related work}

Existing works \cite{Privacy-PreservingEGG, DeepSecure, CryptoNet, Non-interactive, xie2014crypto} for protecting model privacy have tried to avoid leaking the model information to users by isolating users on the client side and the model on the server side. They also provide different strategies to secure model prediction results between the server and the client. Studies \cite{CryptoNet, xie2014crypto} and DeepSecure \cite{DeepSecure} protect model privacy by sending encrypted results back to the client. However, these methods either modify neural networks by replacing activation functions with polynomial approximations, or transform an original neural network to a Boolean circuit. Thus, the complexity on implementation is increased and the performance of the model might be discounted. Moreover, they take no account of the situation when the server itself is an adversary who could steal the whole model on the server side. Ma \emph{et al.} \cite{Non-interactive} discuss a similar scenario but split the neural network into two shares and place them in two servers. They use both HE and secure two-party computation protocols between two servers to preserve model privacy against attack on the server side. Ma \emph{et al.} \cite{Non-interactive} and Rouhani \emph{et al.} \cite{DeepSecure} focus only on inference of a pre-trained model. Compared to these previous works, our proposed design instead naturally protects model privacy from attack on the server side by decentralizing layers into different servers and hiding the server's physical position. It does not require any modification or transformation of the neural network and is naturally applicable to the training phase.

Other current studies on model privacy protection commonly allow model users on the client side to keep certain layers of neural networks while servers keep the rest \cite{VGSR2018, Dropping-Activation, Private-and-Scalable}. Although their primary target is to ensure the data privacy on the client side, model privacy protection is implicitly involved as the adversary on the server only controls parts of the neural network. Nevertheless, model privacy is still at risk because the adversary can infer the other part of the model by analyzing message flows or occupying them. The split learning \cite{VGSR2018} mentioned the multi-hop configuration, which is similar to our decentralized deployment. However, they neither give detailed methods to deploy such a DL model nor focus on model privacy. Our proposed MixNN deploys layers in different servers and uses the method from mix networks to prevent the adversary from controlling the whole model structure and model parameters.

\section{System model}
\label{gen_inst}
There are two parties in the MixNN, namely the designer and servers. A designer is the one who deploys the DL model and processes his or her private data. The servers hold the model layers and provide computation tools and resources. Layers are distributed on different servers and all layers are constructed as a DL model. We consider an adversary who can control a subset of $n$ layers in the system and its goal is to simulate a model $f'(x)$ which is approximately the same as the initial function $f(x)$. We assume each pair of servers is connected by an authenticated point-to-point channel. An adversary who can launch denial-of-service (DOS) attacks is not included in this paper. We also assume that a designer who deploys his/her own DL model using his/her own private data is honest.
\section{System design for MixNN}
\subsection{Setup}
\label{system setup}
There are $m$ servers running in a pool. The designer can acquire servers' information, such as location, configuration, communication speed, price per an hour and so on. The designer can randomly select $n$ servers for deploying the DL model from $m$ ($m\gg{n}$). Among the $n$ servers, $p$ of them contain actual layers (servers) who perform  DL operations and $r$ are dummy servers, namely $n=p+r$. A way to choose these $p$ servers could be based on servers' historical logs, for example, their crash history and performance. He/she assigns $p$ actual layers in $p$ servers. Remaining dummy servers could perform "obscure" operations, for example, transferring messages, or passing through the activation function, e.g. Rectified Linear Unit (ReLU). The $r$ dummy layers can be randomly distributed among these $p$ actual layers. Adding dummy layers among actual layers decreases the possibility for an adversary to control the actual layers and acquire information for simulating a same DL model. 

$m$ servers should register to an authority (this authority could be distributed). Every server in the pool generates its own key pair $pk_i/sk_i$ where $pk_i$ is its public key and $sk_i$ is its secret key, and the server publishes its $pk_i$ and keeps its $sk_i$ secretly. The $i$th server owns its unique $A_i$ address in the system and $A_u$ stands for a designer's address. 

The designer connects $n$ servers as a fixed cascade. Only the designer in the system can pack IP address in the message. To achieve this, we let the designer send a loop message to itself so that no one in the system can know all physical positions of the $n$ servers (layers).

Besides distributing $r$ dummy servers among $p$ actual servers, we use a similar method called "loop message" proposed in \cite{Loopix}, which a server (layer) in the system can send dummy message to another server (layer) in the system. In this way, an adversary cannot know where the message comes from and what it is for.

\subsection{Training phase}
In the training phase, it contains three phases: model initialization phase, forward propagation phase and backward propagation phase. The model initialization is executed only once at the beginning of the training. We set the training with multiple epochs and each epoch includes several iterations. In every iteration, MixNN proceeds one forward propagation and one backward propagation.

In the MixNN, we set every package with the same length so that an adversary cannot tell which type of this package is. A package includes four segments, ($op, en_m, en_{IP}, padding$). The $op$ denotes which type of operation that a layer carries out. It contains four types of operations, 1). $op=0$ means that the designer initializes every layer in a DL model and every layer has to build its corresponding part of the model, 2). $op=1$ stands for a forward propagation message, 3). $op=2$ denotes a backward propagation message, and 4). $op=3$ indicates a testing operation. The $en_m$ means an encrypted message, The $en_{IP}$ denotes an encrypted IP address, and adding $padding$ segments is used to keep the package in a consistent length. 
\begin{figure}
\centering
\subfigure[A cascade is constructed. The designer distributes layers on different servers and each layer generates its key pairs and its IP address.]{\includegraphics[width=10cm]{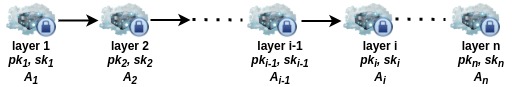}\label{fig_cas}}

\subfigure[Model initialization phase. The designer deploys the size of parameters, parts of the model and optimizer on the corresponding layers.]{\includegraphics[width=10cm]{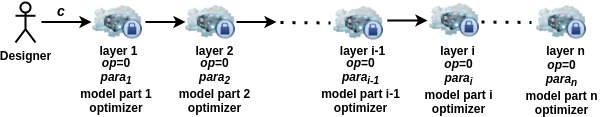}\label{fig_init}}

\subfigure[Forward propagation phase. The designer inputs data into the first layer and the supervised signal into the last layer. Except from the last layer, each layer needs to compute $Z_i$. The last layer should computes the loss $l$. ]{\includegraphics[width=10cm]{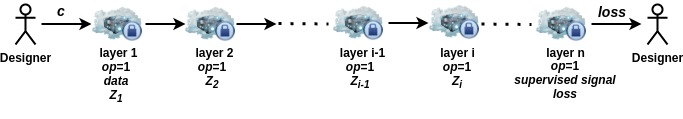}\label{fig_forward}}
\subfigure[Backward propagation phase. For the layer $i$, layer $i$ updates its parameters $\mathcal{W}_{i}$ using intermediate gradient $\frac{\partial{l}}{\partial{z}_{i}}$ from the previous layer, and prepare $\frac{\partial{l}}{\partial{z}_{i-1}}$ for the next layer.] {\includegraphics[width=10cm]{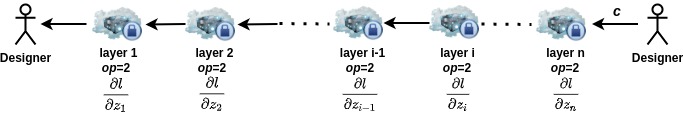}\label{fig_back}}
\caption{Training phase}
\end{figure}

In Fig. \ref{fig_cas}, when a cascade is constructed, only adjacent layers know its previous and next layers' IP address but they have no knowledge about other layers' location. For example, layer 2 only knows the layer 1 and layer 3s' physical IP address and their public keys, but it is hard for layer 2 to acquire layer $i$'s location since the layer $i$'s location is wrapped at the inner of the package. The layer $i$'s location can be acquired in layer $i-1$ assuming there are no failures. For simplicity, we only show the scenario, $n=p$, that is the DL model does not include dummy layers in it.

\subsubsection{Model initialization phase}
In the model initialization phase, the designer needs to distribute the DL model to $p$ servers. For the layers which does DL operations, the designer wraps the operation type, size of parameters and IP address in a message. The designer packs the message as below and send $c$ to the first layer.
\begin{gather*}
    c = (c_1,~A_1), \\
    c_{1} = E_{pk_{1}}(op=0,~para_1,~c_{2},~A_2),\\
    c_{2} = E_{pk_{2}}(op=0,~para_2,~c_{i-1},~A_{i-1}),\\
    \dots\\
    c_{i-1} = E_{pk_{i-1}}(op=0,~para_{i-1},~c_{i},~A_i),\\
    c_i = E_{pk_i}(op=0,~para_i,~c_n,~A_n),\\
    \dots\\
    c_n = E_{pk_n}(op=0,~para_n).
\end{gather*}
The $E_{pk_i}(data)$ means that the encryption algorithm $E$ uses public key $pk_i$ to encrypt data and the ciphertext can only be decrypted by corresponding secret key $sk_i$ with decryption algorithm $D$ ($D_{sk_i}(E_{pk_i}(data))=data$). The ~$para_i$ is the size of parameters, and the $A_i$ is $i$th layer's address. 

After packing the message above, the designer sends $c$ to the layer 1 according to the layer 1's IP address $A_1$. The layer 1 can use its secret key to decrypt the message received from the designer, get the operation type $op=0$, the size of parameter $para_1$, a ciphertext $c_2$, and the layer 2's address $A_2$. Layer 1 builds its corresponding part of the model and optimizer, and sets its own input size of parameter with $para_1$. The parameters of the optimizer like, learning rate and momentum, are the same for all actual layers. We can also send these parameters but it is unnecessary here. The layer 2 to layer $n$ operate the same model initialization as layer 1 does. After finishing the model initialization phase, each layer in a DL model is like what is shown in Fig. \ref{fig_init}.

\subsubsection{Forward propagation phase}
The forward propagation phase is similar to the  initialization phase. However, both forward propagation and backward propagation should be iteratively executed within multiple epochs. Before transmitting the message to the next layer, the layer $i$ needs to compute $\mathcal{Z}_i$, encrypts it with the next layer's public key, and packs it with $c_i$. The designer packs the forward propagation message as below and sends $c$ to the first layer.

\begin{gather*}
        c=(E_{pk_1}(data),~c_1,~A_1),\\
       c_{1} = E_{pk_{1}}(op=1,~c_{2},~A_2),\\
       c_{2} = E_{pk_{2}}(op=1,~c_{i-1},~A_{i-1}),\\
       \dots\\
       c_{i-1} = E_{pk_{i-1}}(op=1,~c_{i},~A_i),\\
\end{gather*}
\begin{gather*}
        c_i = E_{pk_i}(op=1,~c_n,~A_n),\\
        \dots\\
        c_n = E_{pk_n}(op=1,~supervised\ signals,~A_u).
\end{gather*}
Different from other phases in training, we can see that the designer should pack the data as well as the supervised signal, and send the package to the layer 1. The data privacy is not the core part in our paper but we discuss some methods to protect data privacy in the Sec. \ref{Sec_discussion}.

When the layer 1 receives $c$ from the designer, it decrypts the ciphertext and obtains the data, operation type $op=1$, ciphertext $c_1$, and address $A_2$. Layer 1 inputs data to the DL operation in this layer to compute the result $\mathcal{Z}_1$. Then, Layer 1 uses layer 2's public key $pk_2$ to encrypt the $\mathcal{Z}_1$, packs it with $c_1$ and sends ($E_{pk_2}(\mathcal{Z}_1), c_2$) to layer 2. Layer 2 to layer $n-1$ repeat the same steps as what layer 1 does, for example, after decrypting the ciphertext from layer $i-2$,  layer $i-1$ computes $\mathcal{Z}_{i-1}$ and sends ciphertext ($E_{pk_i}(\mathcal{Z}_{i-1}), c_{i}$) to layer $i$. 

Layer $n$ calculates the training loss using $\mathcal{Z}_{n-1}$ and the supervised signal, then encrypts the loss $l$ and sends it back to the designer. The supervised signal is visible only at the last actual layer as it is the most inner part of the package. The designer can also hold the loss layer and supervised signals by him or herself, and hence, supervised signals are protected if it is required. The forward propagation phase is shown in Fig. \ref{fig_forward}.

\subsubsection{Backward propagation phase}
The backward propagation instead starts from layer $n$ to layer 1 which is different from the above two phases. We here only consider gradient descent based methods in updating the DL model parameter. Layer $i$ receives the intermediate gradient $\frac{\partial{l}}{\partial{z}_{i}}$ computed in layer $i+1$, calculates the intermediate gradient $\frac{\partial{l}}{\partial{z}_{i-1}}$ for next layer $i-1$ using chain rule $\frac{\partial{l}}{\partial{z}_{i}}(\frac{\partial{z}_{i}}{\partial{z}_{i-1}})$, and sends it to the next layer. The parameters $\mathcal{W}_i$ in the layer $i$ is updated by first applying chain rule $\frac{\partial{l}}{\partial{z}_{i}}(\frac{\partial{z}_{i}}{\partial{w}_{i}})$, then performing gradient descent related operations. In this phase, the designer only packs the IP address for communication as below. MixNN does not pack any other information (\emph{e.g.}, data or supervised signals) in this phase which is different from the other phases. Finally, the designer sends $c$ to the layer $n$.

\begin{gather*}
    c = (c_n,~A_n),\\
    c_{n} = E_{pk_{n}}(op=2,~c_{i},~A_{i-1}),\\
    \dots\\
    c_{i} = E_{pk_{i}}(op=2,~c_{i-1},~A_{i-1}),\\
    c_{i-1} = E_{pk_{i-1}}(op=2,~c_{2},~A_2),\\
    \dots\\
    c_2 = E_{pk_2}(op=2,~c_1,~A_1),\\
    c_1 = E_{pk_1}(op=2,~A_u). \\
\end{gather*}

Layer $i$ receives message from layer $i+1$, it then decrypts the ciphertext and gets the operation type $op=2$,  intermediate gradient $\frac{\partial{l}}{\partial{z}_{i}}$, ciphertext $c_i$, and address $A_{i-1}$. Layer $i$ calculates $\frac{\partial{l}}{\partial{z}_{i-1}}$ and encrypts it with $pk_{i-1}$. Finally, layer $i$ packs ($E_{pk_{i-1}}(\frac{\partial{l}}{\partial{z}_{i-1}}), c_{i-1}$) and sends it to layer $i-1$. The backward propagation is shown in Fig. \ref{fig_back}.

\subsection{Testing phase}

When the training phase is finished, the designer can perform testing or inference using his or her own metric. The procedure is pretty similar to the forward propagation phase while an input to the metric is needed other than a loss from the model. The packing message is shown below. The testing is only a one-way process, the designer sets the operation type with $op=3$ and decides an ending layer to generate the corresponding input. After one forward propagation, MixNN sends it back to the designer.


\begin{gather*}
        c=(E_{pk_1}(data),~c_1,~A_1),\\
        c_{1} = E_{pk_{1}}(op=3,~c_{2},~A_2),\\
        c_{2} = E_{pk_{2}}(op=3,~c_{i-1},~A_{i-1}),\\
        \dots\\
        c_{i-1} = E_{pk_{i-1}}(op=3,~c_{i},~A_i),\\
        c_i = E_{pk_i}(op=3,~c_n,~A_n),\\
        \dots\\
        c_n = E_{pk_n}(op=3,~A_u).
\end{gather*}

 Take the classification task using probability as an instance, the designer can let the layer $n-1$ send the output of softmax function back to the client and use it to judge the classification performance using various metrics like the confusion matrix, precision, recall and F1 score. 
\section{Security analysis}
We explain how MixNN resists following attacks.
\subsection{Crash failure}
A layer on a server may crash. This degrades the performance of MixNN, especially when the crash occurs in the training phase. We use the following way to defend against this attack. We define $t$ as the maximum communication time when transferring a message between two servers, and $\delta$ as the average time which a server proceeds a message. We denote $n$ as the total number of servers (layers).
\begin{itemize}
    \item A designer sets a time bound $T$ ($T >> n\delta + (n-1)t $) when he or she sends the message to the first server or the last server.
    \item If the designer does not receive the response within time $T$, the designer realizes that the crash failure occurs.
    \item The designer cannot locate crashed servers. A simple way is that the designer replaces all servers in MixNN with other $n$ servers.
\end{itemize}
An another way is that when a server in the cascade does not receive the response from its adjacent server, the server can report the failure to the designer. There are two scenarios here: 1). an honest server reports this failure, 2). a malicious server reports this to achieve its goal like decreasing the credits of an honest server. In MixNN, the designer cannot distinguish the two scenarios, and the simplest way is to replace both the servers. MixNN can use the same approach proposed by Hemi \emph{et al.} \cite{leibowitz2019no} to isolate malicious servers before a cascade transfers the real message. 

\subsection{Byzantine failure}
An adversary dominates a server in a cascade. It can acquire one layer's structure and parameters in a model. Besides that, the adversary can also modify, add or delete the real message \cite{fang2020local} which should be transferred to other layers. Thus, the correctness of a model is affected.

In MixNN, a designer cannot verify each layer's input and output for locating the faulty layer. Meanwhile, verifying each layer's result in every iteration needs more time. In order to guarantee the correctness of the model, we use a simple method that a designer verifies the performance of the model in the testing phase. If he/she finds any problems in that phase, the designer should replace the current $n$ layers with new ones.

In our future work, we will consider whether non-interactive zero knowledge proof \cite{kwon2017atom, tyagi2017stadium} can be used to verify each layer's input and output.

\subsection{Model privacy}
\textbf{Theorem 1}. The MixNN  satisfies the security definition of model privacy.

\textbf{Proof}. The definition of model privacy requires that the adversary $\mathcal{A}$ cannot simulate a model $f'(x)$ which is approximately the same as the initial model $f(x)$. In the adversary model, we have two assumptions, 1). the designer side is honest, 2). an adversary $\mathcal{A}$  control most of layers in a DL model. 

For assumption 1, we assume that an adversary $\mathcal{A}$ cannot acquire the private data, the way that a designer configures layers in a DL model, the number of layers in a DL model, and the construction of cascade (the detailed physical address of these layers) on the designer side. For assumption 2, we assume that an adversary $\mathcal{A}$ cannot control most of layers in a DL model and the adversary $\mathcal{A}$ cannot control both the first layer and the last layer. There is no restriction about how many layers are faulty, for examples, $1/3$ or $1/2$ of total layers. Under these assumptions, we prove that our design satisfies the model privacy in the training phase. We do not consider the model privacy in the testing phase where an adversary can query the model.

\textit{We first consider that an adversary $\mathcal{A}$ controls one layer $i$ ($i \in {1,\cdots,n}$) in a DL model.} 
We assume that operations with parameters are in $\sigma(W_{i}z_{i-1}+b_{i})$ format among all layers, where the $\sigma$ represents the nonlinearity. In the training phase, the adversary $\mathcal{A}$ can acquire the input $\mathcal{Z}_{i-1}$ and output $\mathcal{Z}_{i}$ of layer $i$, intermediate gradient $\frac{\partial{l}}{\partial{z}_{i}}$, where $l$ represents the training loss and ${z}_{i}\in\mathcal{Z}_i$, and intermediate gradient $\frac{\partial{l}}{\partial{z}_{i-1}}$. Then, the $\mathcal{A}$ is able to know the $W_{i}$ and $b_{i}$, the number of rows in $W_{i-1}$, the dimension of $b_{i-1}$, and the number of columns in $W_{i+1}$. Even though the adversary $\mathcal{A}$ can acquire above information from the layer $i$, it is hard for him/her to infer the other layers' structures and parameters with our design. When the number of layers increases, the probability of the adversary $\mathcal{A}$ simulating a $f'(x)$ is negligible.

\textit{We then focus on the situation in which $f$ layers are occupied by the adversary $\mathcal{A}$.}
 We again assume that all layers with parameters have the same type of operations mentioned above. There are two cases below.

\textbf{Case \romannumeral1)} The adversary $\mathcal{A}$ does not know the position of layers in a cascade.

Apparently, to the adversary $\mathcal{A}$, $f$ layers are distributed randomly in this case. The $\mathcal{A}$ indeed knows the parameters in $f$ layers and their adjacent layers' parameter dimension. However, the adversary $\mathcal{A}$ cannot figure out what these layers are and how to combine and construct them as a DL model. When there are more layers in a DL model, knowing these $f$ layers is not much helpful for the adversary $\mathcal{A}$ to simulate an $f'(x)$.

\textbf{Case \romannumeral2)} The adversary $\mathcal{A}$ knows the position of layers in a cascade. 

The most severe attack in this case is shown in Fig. \ref{fig_model}. The adversary $\mathcal{A}$ knows the detailed position of a cascade which constructs a DL Model in the model initialization phase and successfully dominates the layer 2, 4 and 6. In the training phase, the adversary $\mathcal{A}$ knows not only the structures and parameters of 2, 4 and 6 but also the size of parameters in layer 3 and layer 5 and input as well as output of these two layers. Therefore, the adversary $\mathcal{A}$ can find parameters in these two layers by model extraction methods \cite{yan2020cache, StealingMachine}. It means that the adversary $\mathcal{A}$ knows $n-2$ layers between the first layer and the last layer.

However, the model privacy is kept by the first layer and the last layer. The raw data and loss are preserved secretly, hence the adversary $\mathcal{A}$ cannot obtain them or use them to simulate an $f'(x)$. Without the loss layer, the adversary cannot know this model is for. 

This completes the proof of the theorem 1. \hfill $\Box$

\begin{figure*}
  \centering
  \includegraphics[scale=0.5]{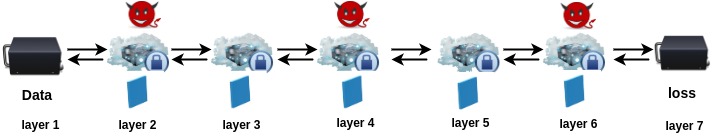}
  \caption{The most severe attack in the second case in MixNN.}
  \label{fig_model}
\end{figure*}

\subsection{Model privacy}
\label{medel_privacy}

Although the MixNN decentralizes and hides model related information so that an adversary cannot fully obtain them for simulating an approximate DL model, a more serious case is that the adversary can successfully occupy the first and last layer which perform DL operations in the current structure. As the input to the model on the first layer and loss as well as supervised signals on the last layer are exposed, the adversary can easily simulate an approximate model using these information. 

In order to avoid this case, the designer can keep one of these two layers or both of them on his or her own hands. Therefore, the adversary can not fully acquire the input, loss or supervised signals.

\subsection{Data privacy}
Federated learning \cite{federatedLearning} and split learning \cite{VGSR2018} facilitate distributed collaborative learning without disclosing original training data. In MixNN, the designer should input the data to the first layer in the forward propagation phase and testing phase. If the first layer is unfortunately controlled by an adversary, the data can be accessed by that adversary. Therefore, the data privacy is not preserved. In order to solve this issue, we provide following methods.

\textit{Distribute the first layer on the client side}. This method is similar to the work mentioned in \cite{privacyOnelayer}, and it avoids the data to be leaked to an adversary. However, when the training is done, how to let other clients use this model is a problem since the first layer is on the designer side. If other clients want to test their dataset, they still need to transmit their data to the designer. A way to solve this is to use obfuscation \cite{obfuscation} to obfuscate the first layer. After the training of a DL model is done, the designer can upload the obfuscated layer to his/her private cloud. Only authorized clients can access it, download it to his/her local machine, and use this part of code as the entrance to the model. 

\textit{Trusted  Execution  Environments  (TEE)}. Another technique to protect confidentiality is using TEE like Intel SGX \cite{SGXcon} or ARM TrustZone \cite{Trustzone}. 
For example, SGX helps to increase protections for sensitive data even when an attacker has full control of the platform. The designer can send encrypted data into the SGX enclave and only the enclave can decrypt the data. However, accelerators such as GPUs do not support TEE, and the SGX has a limited memory size. 

\textit{Homomorphic encryption scheme}. Fully Homomorphic Encryption (FHE) \cite{FHEncryption} is a new class of encryption scheme that allows computing on encrypted data without decryption. FHE has been shown to be useful in many privacy preserving applications such as image classification \cite{CryptoNet, FasterCryptoNets}. In the MixNN, we can use the same methods above but change operations in each layer for processing the encrypted data. However, FHE suffers from two main problems: 1). high computational over-head, and 2). limited arithmetic set (only addition and multiplication on encrypted data are naturally supported).

\subsection{Improvement for the design of MixNN}

\subsubsection{Another configuration with MixNN design}
We only set one layer on one server in the current setting. For DL models with much more layers than the case in the experiment, we do not want too much degradation on the running time. One configuration is that we can randomly compose some adjacent layers on one server. Our next step is to test VGG 16 \cite{simonyan2014very} with this configuration.

\subsubsection{Implementation}
In the current implementation, every layer serves as both a server and a client. As a server, this layer is bound with an IP address and a port number, and it is listening to this channel via (IP, port) and waiting for the information. We do not use multi-threading method to implement it. Hence, when the communication is frequent and the request buffer is full to this layer, it needs to wait for previous requests being processed, then it can settle down other messages. This is another reason why running time becomes longer.

\section{Evaluation}
\label{evaluation}
\subsection{Experiment settings}
\label{expsetup}
We compare the performance and efficiency of a neural network with the same one using MixNN design in the MNIST handwritten digits classification task. Total 30k training digits and 10k test digits are used for training and inference. The performance is defined as the classification accuracy, which is the proportion of correct predictions of the test dataset. The efficiency is measured using running time. 

We employ a multilayer perceptron (MLP) with each layer's configuration listed in Table \ref{table:MLPconfiginMixNN}. It is trained using negative log-likelihood (NLL) loss in server (layer) 5 with logarithm of probabilities (LogSoftmax) from server (layer) 4. We intentionally set server (layer) 4 and 5 with no parameter to simulate a more flexible situation, as our MixNN allows designers to further split or merge operations in different layers of a DL model. The optimization method is the stochastic gradient descent (SGD) with mini-batch size 64, learning rate 0.01 and momentum 0.9.
We use Pytorch to implement these settings.

\begin{table}[!htbp]
  \caption{The configuration of MLP in decentralized servers}
  \label{table:MLPconfiginMixNN}
  \centering
  \begin{tabular}{cccc}
    \toprule
    
    Server Index & Operations  & Input dimension & Output dimension \\
    \midrule
    1          & Linear+ReLU & 784             & 128              \\
    2          & Linear+ReLU & 128             & 64               \\
    3          & Linear      & 64              & 10               \\
    4          & LogSoftmax  & 10              & 10               \\
    5          & NLLloss     & 10              & 1                \\
    \bottomrule
  \end{tabular}
\end{table}

The entire MixNN library is written using Python language. We use Python pycrypto as our crypto library and the public key encryption scheme is RSA with 2048 key length. We deploy MixNN on Amazon AWS. Each instance is run in Ubuntu 16.04 version 43.0 for deep learning. We use $t2.micro$ with one vCPUs and 1GB memory. We run all instances in the same region (Virginia). 

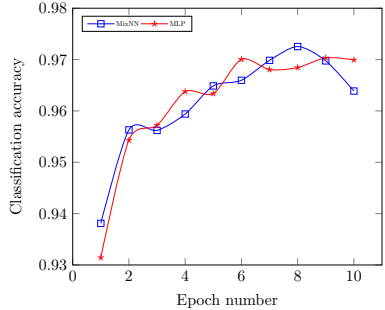
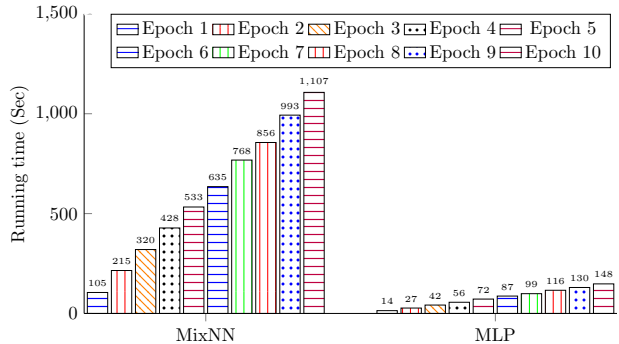
\begin{figure*}[!ht]
\centering
     \subfigure[Classification accuracy compared Mix-NN with MLP.]{
\begin{tikzpicture}[scale=0.6]
  \begin{axis}[
	xlabel={Epoch number},
	ylabel={Classification accuracy},
    legend columns=3,
	legend pos=north west,
    legend cell align=left,
    legend style={
        nodes={scale=0.45, transform shape}},
     ymin=0.93,ymax=0.98,smooth,
     xmin = 0,
    ]
    \addplot [blue,mark=square] coordinates {(1, 0.9380971337579618) (2,0.9563097133757962) (3, 0.9562101910828026) (4, 0.9593949044585988) (5, 0.9648686305732485) (6, 0.9659633757961783) (7, 0.96984474522293) (8, 0.972531847133758) (9, 0.9697452229299363) (10, 0.9638734076433121) };
    
     \addplot [red,mark=star] coordinates {(1, 0.9314291401273885) (2,0.9543192675159236) (3, 0.9572054140127388) (4, 0.9637738853503185) (5, 0.9633757961783439) (6, 0.9700437898089171) (7, 0.9680533439490446) (8, 0.9684514331210191) (9, 0.9703423566878981) (10,0.9699442675159236) };

     \legend{MixNN, MLP}
  \end{axis}
\end{tikzpicture}
		\label{accuracy}
	}
	\hspace{1pt}
	 \subfigure[Running time compared MixNN with MLP.]{
\begin{tikzpicture}[scale=0.7]
\begin{axis}[
    every axis plot post/.style={/pgf/number format/fixed},
    ybar,
    bar width=11pt,
    x=5.5cm,
    ymin=0,
    axis on top,
    ylabel={Running time (Sec)},
    ymax=1500,
    xtick=data,
    enlarge x limits=0.42,
    symbolic x coords={MixNN,MLP},
    visualization depends on=rawy\as\rawy, 
    after end axis/.code={ 
            \draw [ultra thick, white, decoration={snake, amplitude=1pt}, decorate] (rel axis cs:0,1.0) -- (rel axis cs:1,1.0);
        },
    nodes near coords={%
            \pgfmathprintnumber{\rawy}
        },
        every node near coord/.append style={font=\tiny},
    axis lines*=left,
    clip=false,
    legend columns=5,
    legend image code/.code={%
      \draw[#1] (0cm,-0.1cm) rectangle (0.4cm,0.1cm);
    },
    legend style={at={(0.05,1)},anchor=north west}
    ]
\addplot[pattern = horizontal lines, pattern color=blue] coordinates {(MixNN,105) (MLP,14)};

\addplot[pattern = vertical lines, pattern color=red] coordinates {(MixNN,215) (MLP,27)};

\addplot[pattern=north west lines, pattern color=orange] coordinates {(MixNN,320) (MLP,42) };

\addplot[pattern=dots, pattern color=black] coordinates {(MixNN,428) (MLP,56)};

\addplot[pattern=horizontal lines, pattern color=purple] coordinates {(MixNN,533) (MLP,72)};

\addplot[pattern = horizontal lines, pattern color=blue] coordinates {(MixNN,635) (MLP,87)};

\addplot[pattern = vertical lines, pattern color=green] coordinates {(MixNN,768) (MLP,99)};

\addplot[pattern = vertical lines, pattern color=red] coordinates {(MixNN,856) (MLP,116)};

\addplot[pattern=dots, pattern color=blue] coordinates {(MixNN,993) (MLP,130)};

\addplot[pattern=horizontal lines, pattern color=purple] coordinates {(MixNN,1107) (MLP,148)};

\legend{Epoch 1, Epoch 2, Epoch 3, Epoch 4, Epoch 5, Epoch 6, Epoch 7, Epoch 8, Epoch 9, Epoch 10}
\end{axis}
\end{tikzpicture}
		\label{latency}
	}
	
\caption{Classification accuracy and running time compared MixNN with MLP}
	\label{fig:throughputsamen}
\end{figure*}

\subsection{Results and Analysis}
We show the classification accuracy and running time of training with different epochs. We name MLP and MixNN for two different settings in the results for simplicity. In Fig. \ref{accuracy}, we can observe that the differences between classification accuracy of MixNN and MLP are always less than 0.001 in each epoch, thus our MixNN keeps almost same performance in MLP in this task. The reason is obvious as MixNN does not modify the MLP during the training or inference, and we only have different parameter initialization and data shuffling in two settings.

In Fig. \ref{latency}, we can see that running time of MixNN of each epoch is always higher than its counterparts in MLP, and it is 7.5 times higher than MLP's case in average. The reason is that MixNN spends more time on transmitting messages between layers (servers) as well as encrypting and decrypting message flow, and the designer side needs to pack the messages twice in an iteration.

\bibliography{mybibfile}

\end{document}